# $Al_4SiC_4$ vibrational properties: Density Functional Theory calculations compared to Raman and Infrared spectroscopy measurements

L. Pedesseau[1,a)], O. Chaix-Pluchery[2], M. Modreanu[3], D. Chaussende[2], E. Sarigiannidou[2], A. Rolland[1], J. Even[1], and O. Durand[1]

[1]*UMR FOTON, CNRS, INSA Rennes, Rennes, F35708, France*

[2]*Université Grenoble-Alpes, CNRS, LMGP, F-38000 Grenoble, France*

[3]*Tyndall National Institute, University College Cork, Lee Maltings, Cork, Ireland*

**Abstract**

$Al_4SiC_4$ is a wide band gap semiconductor with numerous potential technological applications. We report here the first thorough experimental Raman and Infrared (IR) investigation of vibrational properties of $Al_4SiC_4$ single crystals grown by high temperature solution growth method. The experimental results are compared with the full theoretical analysis of vibrational properties based on Density Functional Theory calculations that are revisited here. We have obtained a good agreement between the experimental and calculated Raman phonon modes and this allowed the symmetry assignment of all the measured Raman modes. We have revisited the DFT calculation of the IR active phonon modes and our results for LO-TO splitting indicate a substantial decrease of $\Delta\omega_{LO-TO}$ compared with the previous reported calculation. Moreover, most of the IR modes have been symmetry assigned from the comparison of the experimental IR spectra with the corresponding Raman spectra and the $Al_4SiC_4$ calculated phonon modes.

[a)] Author to whom correspondence should be addressed. Electronic mail: laurent.pedesseau@insa-rennes.fr

## 1. Introduction

The dearth of low cost, abundant and non-toxic materials for competitive and sustainable solutions for next generation of electronic, photonic and solar technologies strongly highlights all Al-Si-C-based ceramic materials. A new promising horizon for manufacturers of aluminum silicon carbide alloys with breakthrough for environmental science looks towards the future of the energy storage, transparent conductive oxides (TCO), aluminum-silicon-carbon-based LED, wide band gap optoelectronics, attractive mechanical properties, and high temperature operation for electronics.

Among materials intended for high power applications and visible optoelectronics, new families of single crystal carbide and nitride semiconductors are now developing, such as the Al-Si-C-based ceramic materials with the general formula $Al_4C_3(SiC)_n$. Among them, $Al_4SiC_4$ is a low cost potential candidate for power electronics due to its highly desirable combination of unique physical properties such as high thermal conductivity and stability at high temperature [1–4] and promising semiconductor properties [5–9].

$Al_4SiC_4$ single crystal platelets, in the mm scale, were grown very recently using a high temperature solution growth method [10]. Their crystal structure was investigated by TEM observations and XRD analysis, and reported to be hexagonal (Space group *P6$_3$mc*), the c-axis being perpendicular to the crystal surface [6,10]. The experimental lattice parameters obtained from the XRD patterns were in very good agreement with those reported earlier for $Al_4SiC_4$ powder [11] and single crystals [12]. The high quality of the crystals was confirmed both in TEM images [10] and by optical microscopy under cross-polarized geometry (see Supplementary Material) [6].

The first UV-Vis-NIR optical normal incidence transmittance measurements performed on these $Al_4SiC_4$ platelets revealed a large absorption edge in the transmittance curves around 2.4 eV, probably indicating an indirect band gap [6,10]. This experimental indirect band gap estimation disagrees with that of 1.05 eV predicted by previous *ab initio* calculations of the electronic band structure and optical properties [5]. Recently, we reported a state of the art theoretical analysis of the structural, electronic and optical properties of $Al_4SiC_4$ carried out with a standard Density Functional Theory (DFT) including many-bodies corrections approach [6] to provide theoretical knowledge and suitable models to analyze the experimental results. The piezoelectric constants were reported for the first time and compared to the wurtzite wide band gap materials. The Born dynamical effective charge tensor was correlated to the local bonding environment of each atom revealing the major role of the C atoms in the crystal structure by connecting the AlC$_4$ and SiC$_4$ distorted tetrahedral entities. The calculation of the electronic band structure combined with our first investigations of the crystal optical properties using phase modulated spectroscopic ellipsometry indicated that Al$_4$SiC$_4$ has indirect and direct band gap energies of about 2.5 and 3.2 eV respectively. These values are similar to the band gap values of typical wurtzite wide band gap semiconductor materials such as ZnO (3.44 eV) [13], ZnS (3.91eV) [14], GaN (3.30 eV) [15], SiC (2.2eV< $E_g^{SiC}$ <3.33eV) [16,17] and AlN (5.4 eV) [15]. The Al$_4$SiC$_4$ crystal structure can be also described as Al$_4$C$_3$-type and hexagonal SiC-type structural units alternately stacked along the [0001] direction. Indeed, a previous study using first-principles calculations to investigate the Al$_4$SiC$_4$ bonding characteristics demonstrated that the mechanical properties of this compound are nearly related to its inherent layered structure and as a consequence its elastic properties are dominated by the Al$_4$C$_3$-type structural units [18].

The present paper reports the first joint experimental and theoretical study of the vibrational properties of Al$_4$SiC$_4$ single crystals. IR and polarized Raman spectra of Al$_4$SiC$_4$ single crystals or powder were measured first to provide reference spectra for future spectroscopic investigations of this compound. Moreover, a thorough analysis of the experimental measurements has been sustained by our DFT calculations and also compared with other DFT calculations reported recently [7,8]. This allowed us to symmetry assign all the measured Al$_4$SiC$_4$ vibrational modes. The LO-TO splitting was also discussed at the zone center. The Al$_4$SiC$_4$ phonon dispersion curves were also compared to those reported previously [8].

**2. Experimental**

Al$_4$SiC$_4$ single crystals were grown from high purity silicon and aluminum pieces melted in a graphite crucible; the melt was maintained at high temperature (1800°C) before cooling down at a very low and controlled rate. The synthesis procedure was detailed in a previous paper [10].

The crystal structure of typical mm size Al$_4$SiC$_4$ platelets was investigated by TEM observations and XRD analysis and reported to be hexagonal (Space group *P6$_3$mc*), their c-axis being perpendicular to the crystal surface [6,10]. The experimental lattice parameters obtained from the XRD patterns were a = 0.32812 (±0.00045) nm and c = 2.1704 (±0.0055) nm.

Raman spectra were recorded with a Jobin-Yvon/Horiba LabRam spectrometer equipped with a nitrogen cooled charge-coupled device detector. Experiments were conducted in the micro-Raman mode at room temperature in a backscattering geometry. The 514.5 nm line from an argon ion laser was focused to a 1µm$^2$ spot using a x50 long working distance objective. The incident laser power, measured at the sample surface, was close to 0.75 mW. The instrumental resolution was 2.8 ± 0.2 cm$^{-1}$. To go further in the Raman analysis of Al$_4$SiC$_4$, polarized Raman spectra have been collected on the crystal surface (laser propagation along the c-axis) and on the crystal edge (laser propagation parallel to the crystal surface) to identify all Raman active modes of Al$_4$SiC$_4$. Raman spectra have been obtained for an acquisition time varying between 100, 400 and 900 sec, depending on the polarization configuration (parallel polarization/crystal surface, parallel polarization/crystal edge, and crossed polarizations, respectively).

Fourier transform infrared (FTIR) spectroscopy measurements were performed in transmission mode at room temperature on pellets made of Al$_4$SiC$_4$ powder obtained by crystal grinding dispersed in CsI (1.4 mg/150 mg, resp.), using a Bruker Vertex 70V spectrometer. Spectra were collected in the MIR (4000 - 200 cm$^{-1}$, CsI beamsplitter, CsI/DLaTGS detector, 64 scans) with a resolution of 2 cm$^{-1}$ and in the FIR (100-400 cm$^{-1}$, Si beamsplitter, PE/DLaTGS detector, 32 scans) with a resolution of 4 cm$^{-1}$. The perfect recovering of both MIR and FIR spectra in their common area allowed us to join the spectra together to obtain a continuous spectrum. The sample and optical chambers were maintained under vacuum during measurements which makes higher the signal intensity and avoids moisture damage of the CsI beamsplitter.

Attenuated Total Reflection - Fourier transform infrared (ATR - FTIR) spectroscopy covering MIR to FIR (4000 to 100 cm$^{-1}$) were performed at room temperature on Al$_4$SiC$_4$ single crystals using a Thermo Scientific Nicolet iS50 FTIR equipped with a built-in Diamond ATR, and an automated beamsplitter exchanger. Spectra were collected in the MIR (4000 - 400 cm$^{-1}$, KBr beamsplitter, KBr /DLaTGS detector, 64 scans) with a resolution of 2 cm$^{-1}$ and in the FIR (100-400 cm$^{-1}$, Si beamsplitter, PE/DLaTGS detector, 64 scans) with a resolution of 2 cm$^{-1}$.

## 3. DFT calculations

Density function theory calculations were performed using the plane-wave projector augmented wave (PAW) method as implemented in the VASP code [19–22]. The local density approximation (LDA) was used for the exchange-correlation functional [23,24]. Norm-conserving pseudopotentials were constructed for Al [3s$^2$3p$^1$], Si [3s$^2$3p$^2$] and C [2s$^2$2p$^2$] atoms. Atomic valence configurations were similar to the ones used in previous DFT simulations of the ground state [5,6,18]. A plane-wave basis set with an energy cut-off of 950 eV was used to expand the electronic wave-functions. While a 6x6x1 Monkhorst-Pack grid [25] was used in the recent work of Li *et al.* [7,8], we actually reached a better convergence with a reciprocal space integration of 18x18x3. The crystal structure was relaxed until the forces acting on each atom were smaller than $10^{-6}$eV/Å. The energy convergence was accurately reached in considering the tolerance on the residual potential as the difference between the input and output potentials. The density functional perturbation theory [26–29] (DFPT) was used to calculate the phonon spectra following this path Γ-M-L-A-Γ-K-H-A and including the LO-TO splitting [28].

The Raman susceptibility tensors $\alpha^v$ for a given vibrational eigenmode $v$ have been calculated following standard definitions [30–33]:

$$\alpha_{ij}^v = \sqrt{\Omega} \sum_n \sum_l \left.\frac{\partial \chi_{ij}}{\partial r_l}\right|_n \frac{\xi_{nl}^v}{\sqrt{m_n}}, \begin{cases} n \equiv Al, Si, C \\ i, j, l \equiv x, y, z \end{cases}$$

where $\Omega$ is the volume of the primitive cell, $\chi_{ij}$ is the dielectric susceptibility tensor, $r_l$ and $m_n$ are the displacement along the Cartesian axes and the mass of the atom n, respectively, and $\xi_{nl}^v$ is the normalized eigenmode of the atom n in the vibrational mode $v$. The Raman susceptibility tensors were computed for scattered light polarizations parallel and orthogonal to the incident one and used to calculate Raman modes from the edge ($\vec{k}_i // \vec{k}_f \perp oz$) and the surface ($\vec{k}_i // \vec{k}_f // oz$) of a single crystal (c-axis normal the crystal surface), i being the incident light (or field) and f the scattered light (or field).

## 4. Results and discussion

### 4.1. Vibrational modes at Γ point: Symmetry analysis

Al$_4$SiC$_4$ crystallizes in a hexagonal structure (space group *P6$_3$mc* or *C$_{6v}$*) with two formula units per unit cell. In this crystal structure, there are four Al, one Si, and four C nonequivalent atomic positions. The 18 atoms of the primitive cell give rise to 54 zone center vibrational modes. The distribution of the different atoms in special Wyckoff positions is as followed: they are either in special position 2a (one Al, two C and one Si) or in special position 2b (three Al and two C). The site symmetry is *3m* (or *C$_{3v}$*) for all of them. From the correlation method [34,35], we can identify the species of the site group for each lattice vibration and correlate these species with the *C$_{6v}$* factor group species. After subtraction of the three acoustic modes (A$_1$+E$_1$) of the total representation of the crystal, the irreducible representations associated with the optical modes can be expressed as

$$\Gamma = 8A_1 + 9B_2 + 8E_1 + 9E_2$$

where $A_1$ and $E_1$ modes are polar modes and are both Raman and IR active. $E_2$ modes are nonpolar but Raman active. Finally, $B_2$ modes are silent modes for both Raman and IR. The $B_2$ irreducible representation is defined in the present work according to [34–36]. Thus, 25 Raman lines and 16 IR lines are expected in the spectra.

### 4.2. Symmetry assignment of Raman and IR modes in $Al_4SiC_4$

Polarized Raman spectra of $Al_4SiC_4$ are reported in figures 1(a) and 1(b). Due to the $Al_4SiC_4$ hexagonal structure, the X, Y, Z axes of the laboratory frame were taken with Z parallel to the [0001] direction of the crystals, X and Y in the basal plane, a possibility being X // [10-10] and Y // [12-30]. As illustrated in the figures, our measurements have been performed in the following polarization configurations (Porto's notation) to separate the modes of different symmetry:

$$Z(XX)Z \rightarrow A_1 + E_2$$
$$Z(XY)Z \rightarrow E_2 + \varepsilon A_1$$
$$X(ZZ)X \text{ or } X(YY)X \rightarrow A_1 \text{ or } A_1 + E_2$$
$$X(ZY)X \rightarrow E_1$$

It is to be noted that i) we cannot distinguish between X and Y axes in our crystal platelets, ii) the polarization is not pure in the Z(XY)Z polarization configuration due to the hexagonal structure, this would lead to some (XX) polarization component and thus to the appearance of weak $A_1$ modes ($\varepsilon A_1$ in the corresponding equation).

Spectra collected in the parallel Z(XX)Z and X(ZZ)X polarization configurations (figures 1(a) and 1(b), resp.) allow us to assign with certainty the most intense modes at 865, 543, 393 cm$^{-1}$ as $A_1$ modes. Despite their weak intensity, other modes of $A_1$ symmetry can be assigned, either from the Z(XX)Z polarization configuration at 317, 718 and 919 cm$^{-1}$ (figure 1(a)), or from the X(ZZ)X one at 306 cm$^{-1}$ (figure 1(b)), or from both configurations at 597 cm$^{-1}$. The crossed Z(XY)Z polarization configuration (figure 1(a)) allows to measure mainly the modes of $E_2$ symmetry and additional weak $A_1$ modes as mentioned above. Without any doubt, the modes at 84, 100, 255, 486, 690 and 787 cm$^{-1}$ can be assigned as $E_2$ modes, this is also confirmed by our DFT simulation (Table I). The calculation assistance is more evident for the assignment of some other modes such as the mode at 281 cm$^{-1}$. This mode has finally a $E_2$ symmetry but its assignment was difficult due to the presence of an intense $E_1$ mode at the same position (see X(ZY)X spectrum in figure 1(b)). Nevertheless, it is to be noted that all these seven $E_2$ modes are also visible in the spectrum measured in the parallel Z(XX)Z polarization configuration in addition to the $A_1$ modes, as expected from the selection rules. The spectrum measured in the crossed X(ZY)X polarization configuration allows us to assign the modes at 189, 244, 281, 478 and 686 cm$^{-1}$ as $E_1$ modes. Although $A_1$ modes are forbidden in this polarization configuration, some of them are nonetheless observed in the spectrum.

In summary, we have assigned all the eight expected $A_1$ modes, seven $E_2$ modes (nine modes expected) and five $E_1$ modes (eight modes expected). However, some weak additional modes remain unidentified: two modes are visible in the four spectra at 357-360 and 935 cm$^{-1}$, a third one at 660 cm$^{-1}$ in three spectra, a fourth one at 210 cm$^{-1}$ only in the Z(XX)Z polarization configuration. This means that the first three modes are not polarization dependent and must have other origins. In fact, for the mode at 210 cm$^{-1}$, a possible $E_2$ assignment can be considered as all the $A_1$ modes are assigned.

Absorbance spectra obtained from Diamond ATR and FTIR measurements are reported in figures 2(a) and 2(b); they show not well-defined lines although these spectra are the best which have been measured from our samples. Measurements were made difficult due to the presence of polarization effects inherent in ATR geometry, to the very small size of the crystals and to the strong IR absorbance of $Al_4SiC_4$ in the Mid to Far IR spectral range. While the observed IR modes are weak and not well-resolved, they are in good agreement with the Raman modes discussed and with the theoretical DFT calculations. It is to be noticed that ATR spectra are composed of positive and negative peaks as seen in figure 2(a) and that differences in line positions are generally observed in ATR spectra in comparison with spectra measured with the FTIR technique.

A comparison between the experimental values of $Al_4SiC_4$ phonon wavenumbers obtained by Raman and IR spectroscopy and DFT values calculated at the zone-center by us and by Li *et al.* [7,8] is reported in Table I. An overall agreement is observed for the symmetry assignment of Raman modes in comparison with theoretical modes, including the mode at 210 cm$^{-1}$ which is assigned to $E_2$ symmetry. The positions of a majority of Raman lines are closer to our theoretical values in comparison with values reported in Ref. [8]. This can be related to the grid used, 18x18x3 in our case. Although the correspondence is less easy with the IR lines, most of them have been assigned from the theoretical modes and from the Raman modes as shown in Table I. However, an uncertainty remains about the $A_1$ or $E_1$ symmetry assignment of the IR mode at 612 cm$^{-1}$. Fortunately, the comparison with the ATR spectrum is more favorable to distinguish between the two symmetries because of its likeness with the line at 586 cm$^{-1}$, assigned as $A_1$ mode. As for Raman modes and except for some particular modes, the positions of most of IR modes are not so far from our calculated positions both in ATR spectra (146, 181, 325, 398, 541, 586, 630, 813, 841 cm$^{-1}$) and in FTIR spectra (147, 198, 305, 323, 401, 497, 521 cm$^{-1}$). Others are closer to the calculated positions reported in Ref. [8] (750 cm$^{-1}$ in ATR spectra; 258, 277, 764, 777, 800 cm$^{-1}$ in FTIR spectra). Modes observed at 355-354, 454-457, 890-878 cm$^{-1}$ in ATR-FTIR spectra, respectively, at 664 and 954 cm$^{-1}$ in ATR spectra only, are non-assigned modes and three of them can be related to Raman modes pointed in the same regions; these modes have positions close to the ones calculated for $B_2$ modes, which is quite surprising (Table I). As our crystals are of high quality, no explanation can be given for these observations. However, some anomalous Raman modes have been already reported in several papers dealing with doped- and undoped ZnO grown by different methods. Most of them have been attributed to impurities or defects. In the case of the wurtzite ZnO crystal, Manjón *et al.* [37] considered most of these anomalous modes as silent modes which were disorder-activated due to the relaxation of the Raman selection rules produced by the breakdown of the crystal symmetry induced by defects and impurities. So, for this reason, the three modes can be reasonably assigned to $B_2$ modes which are silent modes in pristine crystal. The last non assigned IR line is observed in the ATR spectrum at 200 cm$^{-1}$. This position appears very close to that of the $E_2$ mode calculated at 207 cm$^{-1}$ in this work, at ~204 cm$^{-1}$ in Ref. [8], and observed at 210 cm$^{-1}$ in the Raman spectrum. However, $E_2$ modes are not IR-active, and as $B_2$ modes, it could come from the presence of defects or impurities in the crystals.

### 4.3. Theoretical study of the phonon dispersion in Al$_4$SiC$_4$

The Al$_4$SiC$_4$ phonon dispersion curves calculated at 0K by DFPT are reported in figure 3. Although the comparison with those obtained in the previous theoretical study [7,8] is rather difficult, both studies agree about the stability of the crystals as indicated by the fact that all phonon frequencies are positive.

The magnitude of the LO-TO splitting, $\Delta\omega_{LO-TO}$ (cm$^{-1}$), calculated for each polar mode, is reported in Table I for small k-points following the path M–Γ–A as well as the corresponding values obtained with the previous theoretical study [7]. The LO-TO splitting, for a given mode, is proportional to the mode oscillator strength and thus, it reflects the intensity of this specific mode. Our calculations reveal that the values corresponding to the largest LO-TO splittings (italic numbers in Table I) are lower than those predicted in the previous study [7] and above all, the IR mode with the maximum splitting is different in both studies: we obtain splitting values of 229.5 cm$^{-1}$ instead of 303.62 cm$^{-1}$ in the previous study for the $E_1$ mode calculated at 803/777.6 cm$^{-1}$, 237 cm$^{-1}$ instead of 363.34 cm$^{-1}$ for the A$_1$ mode calculated at 747/720.1 cm$^{-1}$, 229.5 cm$^{-1}$ instead of 257.99 cm$^{-1}$ for the E$_1$ mode calculated at 483/443 cm$^{-1}$. The maximum value that we found is about 247 cm$^{-1}$ instead of 234.44 cm$^{-1}$ in the previous study [7] for the A$_1$ mode calculated at 844/807.7 cm$^{-1}$.

## 5. Conclusion

Experimental measurements of Raman and IR spectra of Al$_4$SiC$_4$ single crystals are reported for the first time. A theoretical analysis of the vibrational properties of Al$_4$SiC$_4$ using DFPT calculations of the dynamical matrix at the Brillouin zone center is also reported. Despite not well-defined IR lines, a good agreement between experimental and calculated positions of the modes is observed for both Raman and IR spectra, allowing the symmetry assignment of all measured modes. New calculations of the Al$_4$SiC$_4$ phonon dispersion curves have been performed at 0K by DFPT with a better grid of k-points. The LO-TO splitting values have been dramatically decreased compared to those obtained by Li *et al.* [7]. Finally, the order of the largest LO-TO splitting values $\Delta\omega_{LO-TO}$ observed in both studies has been also modified.

## ACKNOWLEDGMENTS

The *ab initio* simulations have been performed on HPC resources of CINES under the allocations 2015-[x2015096724] and 2016-[x2016096724] made by GENCI (Grand Equipement National de Calcul Intensif).

# Figure captions

**Figure 1**: Polarized Raman spectra of $Al_4SiC_4$ single crystals collected in various polarization configurations: (a) spectra collected on the crystal surface (laser propagation along the c-axis), (b) spectra collected on the crystal edge (laser propagation parallel to the crystal surface) ($\lambda$ = 514.5 nm).

**Figure 2**: (a) Diamond ATR spectra of $Al_4SiC_4$ single crystals recorded in the FIR and MIR; (b) FTIR spectra of pellets of $Al_4SiC_4$ powder dispersed in CsI (1.4 mg/150 mg, resp.).

**Figure 3**: $Al_4SiC_4$ phonon dispersion curves calculated by DFT at 0 K.

**Table I**: Calculated ($\omega_{calc}$) and experimental ($\omega_{exp}$) positions, in cm$^{-1}$, of Al$_4$SiC$_4$ Raman and IR modes and calculated values of the LO-TO splitting. Calculated positions from Ref. [8] and calculated values of the LO-TO splitting from Ref. [7] are also reported for comparison. Experimental values were obtained from polarized Raman spectra and ATR spectra of Al$_4$SiC$_4$ single crystals, and from FTIR spectra of pellets made of Al$_4$SiC$_4$ powder dispersed in CsI (1.4 mg/150 mg, resp.). The positions in bold are related to non-assigned modes; the largest LO-TO splitting values appear as italic numbers. Note that $\Delta\omega_{LO-TO} = \sqrt{\omega_{LO}^2 - \omega_{TO}^2}$ is calculated following Γ-M and Γ-A directions.

| | Li *et al* study | | Our study | | | | |
|---|---|---|---|---|---|---|---|
| Symmetry | $\omega_{calc}$[8] | $\Delta\omega_{LO-TO}$[7] | $\omega_{calc}$ | $\Delta\omega_{LO-TO}$ | $\omega_{exp}$ Raman | $\omega_{exp}$ ATR | $\omega_{exp}$ FTIR |
| E$_2$ | 78.0 | | 84 | | 84 | | |
| E$_2$ | 96.0 | | 101 | | 100 | | |
| E$_1$ | 139.7 | 22.39 | 147 | 34 | | 146 | 147 |
| B$_2$ | | | 171 | | | | |
| B$_2$ | | | 177 | | | | |
| E$_1$ | 188.6 | 9.18 | 192 | 8.5 | 189 | 181 | 198 |
| E$_2$ | 203.9 | | 207 | | 210 | **200** | |
| E$_2$ | 262.6 | | 276 | | 255 | | 258 |
| E$_1$ | 268.1 | 9.96 | 280 | 2.5 | 244 | 220 | 233 |
| E$_1$ | 278.2 | 13.03 | 291 | 16 | 281 | | 277 |
| E$_2$ | 280.5 | | 293 | | 281 | | |
| A$_1$ | 290.1 | 73.02 | 293 | 95 | 306 | | 305 |
| A$_1$ | 303.4 | 85.03 | 319 | 69.5 | 317 | 325 | 323 |
| B$_2$ | | | 365 | | **357-360** | **355** | **354** |
| A$_1$ | 397.9 | 2.84 | 409 | 5 | 393 | 398 | 401 |
| B$_2$ | | | 433 | | | **454** | **457** |
| E$_1$ | 443.0 | *257.99* | 483 | *229.5* | 478 | 500 | 497 |
| E$_2$ | 443.6 | | 483 | | 486 | | |
| A$_1$ | 517.8 | 120.06 | 535 | 122.5 | 543 | 541 | 521 |
| B$_2$ | | | 535 | | | | |
| A$_1$ | 578.4 | 245.42 | 593 | 196 | 597 | 586 | 612 |
| E$_1$ | 596.4 | 230.89 | 624 | 189 | 686 | 630 | 612 |
| E$_2$ | 596.8 | | 624 | | 690 | | |
| B$_2$ | | | 652 | | **660** | **664** | |
| A$_1$ | 720.1 | *363.34* | 747 | *237* | 718 | 701 | 714 |
| E$_1$ | 755.5 | 70.13 | 784 | 86 | | 750 | 764 |
| E$_2$ | 755.5 | | 784 | | 787 | | |
| E$_1$ | 777.6 | *363.62* | 803 | *229.5* | | 813 | 777 |
| E$_2$ | 777.7 | | 803 | | | | |
| B$_2$ | | | 835 | | | | |
| A$_1$ | 807.7 | *234.44* | 844 | *247* | 865 | 841 | 800 |
| B$_2$ | | | 880 | | | **890** | **878** |
| A$_1$ | 862.0 | 116.80 | 898 | 166.5 | 919 | 920 | |
| B$_2$ | | | 913 | | **935** | **954** | |

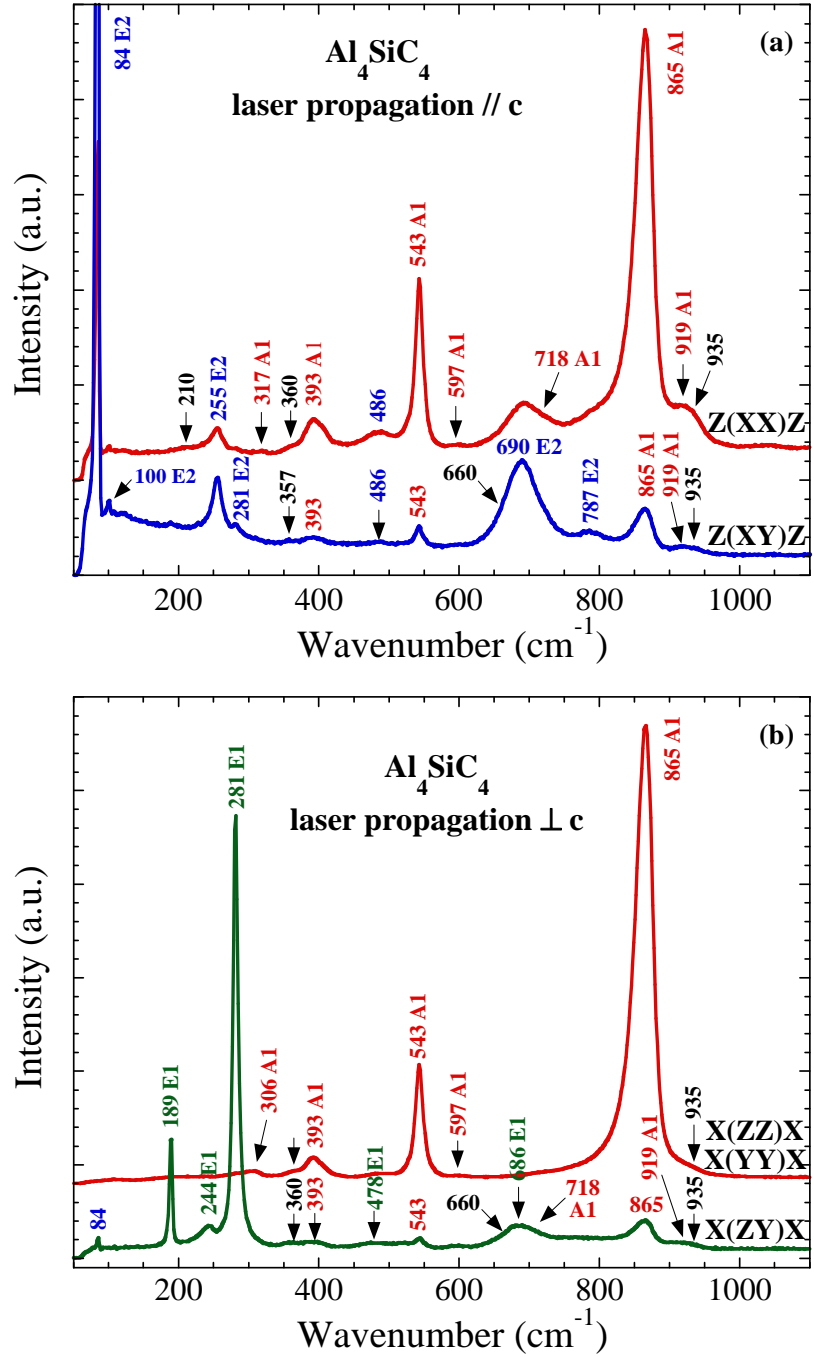

**Figure 1**: Polarized Raman spectra of Al$_4$SiC$_4$ single crystals collected in various polarization configurations: (a) spectra collected on the crystal surface (laser propagation along the c-axis), (b) spectra collected on the crystal edge (laser propagation parallel to the crystal surface) ($\lambda$ = 514.5 nm).

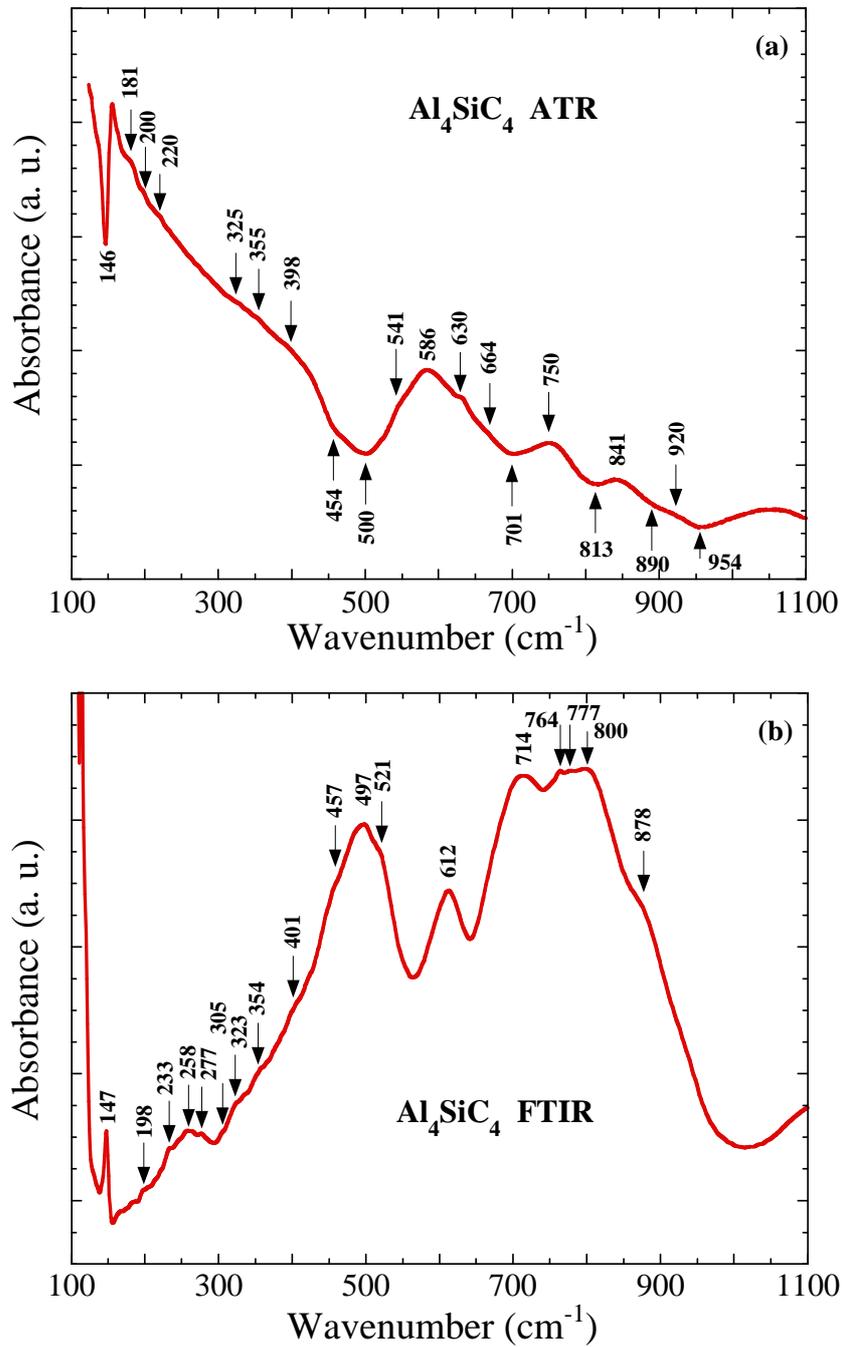

**Figure 2**: (a) Diamond ATR spectra of Al$_4$SiC$_4$ single crystals recorded in the FIR and MIR; (b) FTIR spectra of pellets of Al$_4$SiC$_4$ powder dispersed in CsI (1.4 mg/150 mg, resp.).

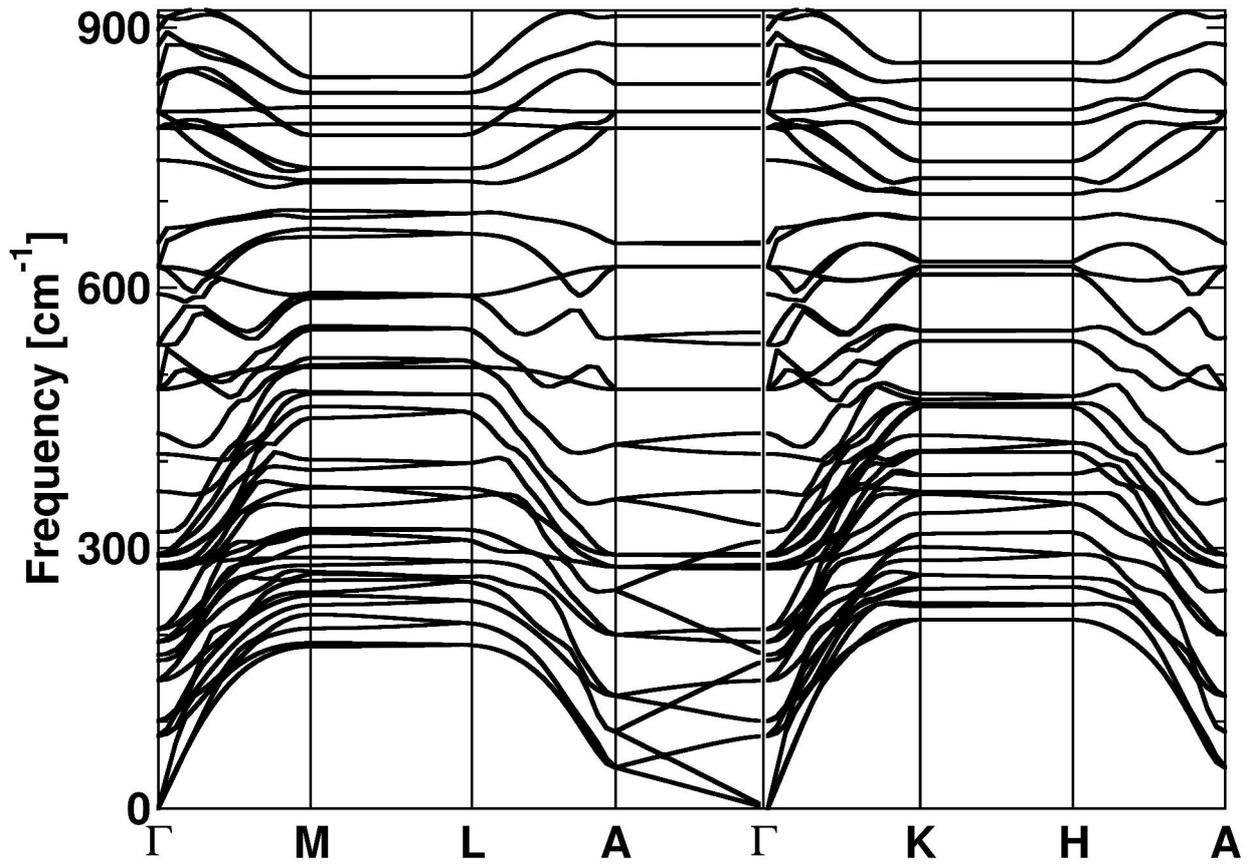

**Figure 3:** Al$_4$SiC$_4$ phonon dispersion curves calculated by DFT at 0 K.